\documentclass[pra,twocolumn]{revtex4-1}
\usepackage{amsmath,amssymb,graphicx}

\begin{document}

\newcommand{\bra}[1]    {\langle #1|}
\newcommand{\ket}[1]    {| #1 \rangle}
\newcommand{\braket}[2]    {\langle #1 | #2 \rangle}
\newcommand{\braketbig}[2]    {\big\langle #1 \big| #2 \big\rangle}
\newcommand{\braketbigg}[2]    {\bigg\langle #1 \bigg| #2 \bigg\rangle}
\newcommand{\tr}[1]    {{\rm Tr}\left[ #1 \right]}
\newcommand{\av}[1]    {\langle{#1}\rangle}
\newcommand{\avbig}[1]    {\big\langle{#1}\big\rangle}
\newcommand{\avbigg}[1]    {\bigg\langle{#1}\bigg\rangle}
\newcommand{\x}{\mathbf{r}}
\newcommand{\bk}{\mathbf{k}}
\newcommand{\bp}{\mathbf{p}}
\newcommand{\re}{\mathrm{Re}}
\newcommand{\im}{\mathrm{Im}}

\title{Quantum interferometry in multi-mode systems}

\author{J. Chwede\'nczuk}
\affiliation{Faculty of Physics, University of Warsaw, ul. Pasteura 5, PL--02--093 Warszawa, Poland}

\begin{abstract}
  We consider the situation when the signal propagating through each arm of an interferometer has a complicated multi-mode structure. 
  We find the relation between the particle-entanglement and the possibility to surpass the shot-noise limit of the phase estimation. Our results are general---they 
  apply to pure and mixed states of identical and distinguishable particles (or combinations of both), for a fixed and fluctuating number of particles. We also show that the 
  method for detecting the entanglement often used in two-mode system can give misleading results when applied to the multi-mode case.
\end{abstract}

\maketitle

\section{Introduction}

Interferometers, the most precise metrological instruments constructed by humans, have played a major role in many break-through experiments. In the famous Michelson-Morely failed attempt to detect
the ether, the device, now called the ``Michelson interferometer'', was used \cite{schumacher1994special}. The negative result of this experiment was relevant for the subsequent formulation 
of the special theory of relativity.
Almost 130 years later, the LIGO team used a similar  Michelson setup to detect the gravitational wave coming from a collision of two black holes, which took place over 1 bilion years ago \cite{grav1}.
At the moment of the detection, the sensitivity of the LIGO instruments was high enough to observe the displacement $\Delta L$ of the test masses $\sim10^{-22}$ times smaller than the length $L$ of the
interferometric arms. With $L\simeq4$\,km, this gives a truly impressive $\Delta L\simeq10^{-19}$\,m, which is $\sim10^3$ times smaller than the radius of a proton. 

Matter-wave interferometers at this
moment can measure the gravitational acceleration $g$ with the sensitivity of $\Delta g\simeq10\frac{\rm nm}{\rm s}$ \cite{gravimetry,ferrari,poli,kas1,kas2,kas3,smallg,smallg2}. 
Once miniaturized, such device could serve as an ultra-precise
geological instrument. An atomic interferometer calibrated through the measurement of the close to the surface Casimir-Polder forces could yield information about the gravitational constant
$G$ and the possible deviations from the Newton $1/r^2$ scaling of the gravitational force at small distances \cite{casimir1,casimir2,casimir3,casimir4,casimir5,G1}. The gravity-field curvature 
has recently been observed \cite{curva} using an atomic interferometer. Interference of the matter waves is also used in the ultra-precise measurements of the fine-structure constant $\alpha$,
which is of fundamental importance \cite{fine1,fine2,fine3,fine4,fine5}.

At the current stage, these devices operate at best at the shot-noise level, i.e., the sensitivity of the estimation of the parameter $\theta$ does not break the shot-noise limit (SNL),
$\Delta\theta\propto\frac1{\sqrt N}$. Here $N$ is the number of probes---for instance atoms or photons---which carry the information about the interferometric phase. However, theoretical results and
multiple proof-of-principle experiments show that the entanglement between these probes is a resource for the
sub shot-noise (SSN) sensitivity \cite{giovannetti2004quantum,pezze2009entanglement}. Those experiments follow different routes to create non-classical states of light or matter.
For instance, the LIGO interferometer already displayed the SSN sensitivity when one of its input ports was fed with a squeezed state of light \cite{ligo2011gravitational}. 
For interferometers operating on matter--waves,   
the non-classicality is associated with the entanglement between the particles. This effect often manifests through the spin-squeezing of the sample 
\cite{esteve2008squeezing,berrada2013integrated,leroux2010orientation,chen2011conditional,gross2010nonlinear,appel2009mesoscopic}.
It is a phenomenon---quantified by the spin-squeeizng parameter \cite{kitagawa1993squeezed,wineland1994squeezed}---associated with the two-mode algebra. The use of this algebra is quite
natural when discussing the interferometric problems---the two arms are identified with the two modes. Most of the interferometric arguments, such as that relating the entanglement to the SSN
sensitivity, are also invoked within this two-mode description. However, in principle, the signal propagating through each arm can have a rich multi-mode structure. This might be a result of
thermal excitations, as witnessed in \cite{esteve2008squeezing}, or the inherently multi-mode nature of the process generating the entangled sample, as in
\cite{lucke2011twin,twin_beam,cauchy_paris,collision_paris,truscott}.

Here we generalize the central theorem of quantum metrology---that relating the SSN sensitivity to the entanglement---to the case when each arm of the interferometer has a complicated 
multi-mode structure \cite{mm1}.
We show that the particle entanglement remains the key resource for beating the SNL. Our proof is of complete generality---it does not make any assumption about the state and works for 
identical and distinguishable
particles or combination of both. Also, it applies to systems with a fixed or fluctuating number of particles. We derive the sensitivity of the phase estimation from the measurement of the population
imbalance between the two arms. Finally, we show how the method of detecting the particle entanglement, which works for the two-mode systems, can incorrectly indicate the presence of non-classical correlations in the multi-mode configurations.

This work is organized as follows. In Section \ref{sec.ent} we derive the relation between the entanglement and the SSN sensitivity for identical (Section \ref{sec.ent.id}) and distinguishable particles
(Section \ref{sec.ent.dist}). Other configurations are discussed in Section \ref{sec.ent.other}. In Section \ref{sec.est} we calculate the phase sensitivity for the estimation based on the
knowledge of the average population imbalance between the two arms. In Section \ref{sec.det} we show that the entanglement witness often used for two-mode systems, can give false results in the multi-mode setups. The summary is contained in Section \ref{sec.sum}.

\section{Entanglement and SSN sensitivity in multi-mode systems}\label{sec.ent}

\subsection{Bosons}\label{sec.ent.id}
\subsubsection{Multi-mode interferometric transformations}

\begin{figure*}[t!]
  \includegraphics[width=\textwidth]{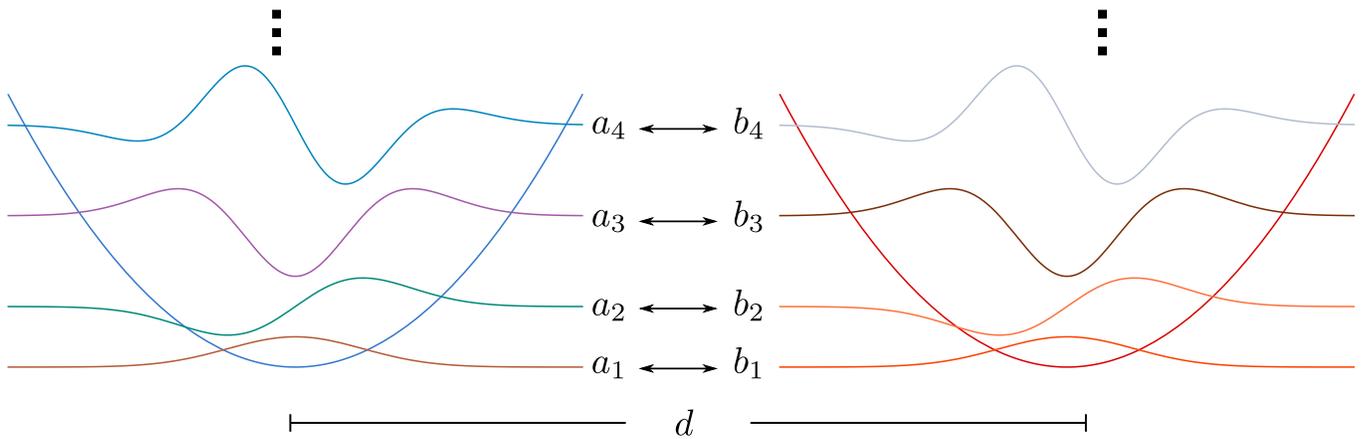}
  \caption{The two-arm, multi-mode interferometer, illustrated here with the eigen-modes of the
    harmonic oscillator. Each arm (here a harmonic well) has the same ladder of states, just shifted by $d$.
    The interferometric transformation can either imprint the relative phase
    between the arms or move the particles from one arm to the other. }
  \label{fig.ladder}
\end{figure*}

Let $a$ and $b$ denote the two arms of an interferometer. Here, for illustration, we will assume that these arms are spatially separated, however it could be the momentum or other degree of freedom, such as the fine structure, that distinguish $a$ from $b$. In the standard two-mode case, with each arm a single operator $\hat a$ and $\hat b$ is associated. 
The interferometric transformations are generated by the angular-momentum operators
\begin{subequations}\label{J_2m}
  \begin{eqnarray}
    &&\hat J_x=\frac12(\hat a^\dagger\hat b+\hat b^\dagger\hat a),\label{x_2m}\\
    &&\hat J_y=\frac1{2i}(\hat a^\dagger\hat b-\hat b^\dagger\hat a).\label{y_2m}\\
    &&\hat J_z=\frac12(\hat a^\dagger\hat a-\hat b^\dagger\hat b).\label{Jz}
  \end{eqnarray}
\end{subequations}
To account for the multi-mode structure of each arm, we introduce the bosonic field operator $\hat\Psi(\x)$ which consists of two parts, i.e.,
\begin{equation}\label{field}
  \hat\Psi(\x)=\hat\Psi_a(\x)+\hat\Psi_b(\x).
\end{equation}
Our aim is to construct the interferometric transformation
which will act on the multi-mode arms $a$ and $b$, rather than on two modes only. It is clear, that such operations can be constructed in many different ways 
and that the interferometer's performance depends on the structure of
the regions as well as on our choice of transformations. 
To limit the number of possibilities, we will first assume that the mode structure of $a$ and $b$
is the same, and it is only the spatial separation that makes the distinction between them, see Fig.~\ref{fig.ladder}. 
Therefore, each operator can be expanded into its corresponding basis
\begin{subequations}\label{ops}
  \begin{eqnarray}
    &&\hat\Psi_{a}(\x)=\sum_{n}\psi^{(n)}_{a}(\x)\hat a_n,\\
    &&\hat\Psi_b(\x)=\sum_{n}\psi^{(n)}_b(\x)\hat b_n,
  \end{eqnarray}
\end{subequations}
and the spatial wave-functions are shifted by $\mathbf d$, i.e.,
\begin{equation}\label{rel1}
  \psi^{(n)}_b(\x+\mathbf{d})=\psi^{(n)}_a(\x).
\end{equation}
Also, we assume $|\mathbf d|$ to be much larger than the characteristic widths of the wave-packets, giving
\begin{equation}\label{rel2}
  \psi^{(n)}_{a}(\x)\psi^{(m)}_{b}(\x)=0\ \ \   \forall\ n,m.
\end{equation}
Once the spatial structure of the system is determined, we pick the interferometric transformations. To establish the analogy with the two-mode case, we
consider two types of transformations---the mode mixing and the phase imprint. A natural extension of (\ref{J_2m}) is
\begin{subequations}\label{J_mm}
  \begin{eqnarray}
    &&\hat J_x=\frac12\int\!\! d\x(\hat\Psi^\dagger_a(\x)\hat\Psi_b(\x+\mathbf{d})+\hat\Psi^\dagger_b(\x+\mathbf{d})\hat\Psi_a(\x))\\
    &&\hat J_y=\frac1{2i}\int\!\! d\x(\hat\Psi^\dagger_a(\x)\hat\Psi_b(\x+\mathbf{d})-\hat\Psi^\dagger_b(\x+\mathbf{d})\hat\Psi_a(\x))\\
    &&\hat J_z=\frac1{2}\int\!\! d\x(\hat\Psi^\dagger_a(\x)\hat\Psi_a(\x)-\hat\Psi^\dagger_b(\x)\hat\Psi_b(\x)).\label{mJz}
  \end{eqnarray}
\end{subequations}
These integrals can be calculated using relations (\ref{rel1}) and (\ref{rel2}), and the outcome is
\begin{subequations}\label{mJ}
  \begin{eqnarray}
    &&\hat J_x=\frac12\sum_n(\hat a_n^\dagger\hat b_n+\hat b_n^\dagger\hat a_n)\equiv\sum_n\hat J_x^{(n)}\\
    &&\hat J_y=\frac1{2i}\sum_n(\hat a_n^\dagger\hat b_n-\hat b_n^\dagger\hat a_n)\equiv\sum_n\hat J_y^{(n)}\label{y_mm}\\
    &&\hat J_z=\frac12\sum_n(\hat a_n^\dagger\hat a_n-\hat b^\dagger_n\hat b_n)\equiv\sum_n\hat J_z^{(n)}.
  \end{eqnarray}
\end{subequations}
Therefore, the mode mixing operators act on each pair of modes separately. This is a result of 1$^\circ$ the assumption about the symmetry between the regions and 2$^\circ$
the particular form of the coupling
in Equations (\ref{J_mm}). While 1$^\circ$ can be regarded as unnatural or highly idealistic, such symmetry is encountered for instance in twin-beam configurations formed by the scattering
of atoms from a Bose-Einstein condensate \cite{wasak_twin}. On the other hand, 2$^\circ$ seems a natural choice: the interferometer should simply ``copy'' a particle from one arm to another.
This second condition can be expressed in other words: the interferometric transformation does not use any knowledge about the internal structure of the arms. It treats each arm as a 
whole and does not penetrate the inner structure.

The multi-mode interferometric transformations which are single-particle operations (i.e., do not entangle the resources) are usually considered in a form
\begin{equation}\label{trans}
  \hat U(\theta)=e^{-i\theta\hat J_n}.
\end{equation}
Here $\hat J_n=\vec n\cdot\hat{\vec J}$ is a scalar product of a unit vector $\vec n$ and a vector of the 
angular momentum operators~\footnote{Note that in principle, an interferometer can be a series of linear transformations, rather than a single one \cite{chwedenczuk2010rabi} 
  Still, such composite operation remains linear and does not entangle the particles.}. We will now demonstrate that,
in analogy to the two-mode case, the particle entanglement is a necessary resource for beating the shot-noise limit of the phase estimation also in the multi-mode case.

\subsubsection{Role of the particle entanglement}\label{sec.qfi}

Let us begin with a two-mode separable (i.e., non-entangled) pure state of $N$ bosons. It is the spin-coherent state
\begin{equation}\label{css_2m}
  \ket{z,\varphi;N}=\frac1{\sqrt{N!}}\left(\sqrt ze^{i\varphi}\hat a^\dagger+\sqrt{1-z}\hat b^\dagger\right)^N\ket0.
\end{equation}
Here, $z\in[0,1]$ is the population imbalance between the two modes, while $\varphi\in[0,2\pi]$ is the relative phase. This state is a basic building block of the density matrix of $N$
non-entangled bosons, which reads
\begin{equation}\label{sep_2m}
  \hat\varrho_{\rm sep}=\int_0^1\!\!dz\int_0^{2\pi}\!\!\!d\varphi\,{\mathcal P}(z,\varphi)\ket{z,\varphi;N}\bra{z,\varphi;N},
\end{equation}
where ${\mathcal P}$ is a probability distribution of the variables $(z,\varphi)$. The expression (\ref{sep_2m}) is the fixed-$N$ analog of the classical state of the electromagnetic field,
according to the Glauber-Sudarshan criterion \cite{glaub,sud,cauchy,cauchy_long}.
The expressions (\ref{css_2m}) and (\ref{sep_2m}) can be easily generalized to the multi-mode setup. Namely, the former transforms to
\begin{equation}\label{css_mm}
  \ket{z,\varphi;N}\rightarrow\ket{\vec\alpha,\vec\beta;N}=\frac1{\sqrt{N!}}\left(\vec\alpha\,\hat{\vec a}^\dagger+\vec\beta\,\hat{\vec b}^\dagger\right)^N\ket0.
\end{equation}
The vectors of complex amplitudes $\vec\alpha$ and $\vec\beta$ are normalized, i.e, $|\vec\alpha|^2+|\vec\beta|^2=1$, while
$\hat{\vec a}$  and $\hat{\vec b}$ are vectors of mode operators introduced in Eq.~(\ref{ops}). Similarly, the density matrix of unentangled bosons now reads \cite{wasak2016role}
\begin{equation}\label{sep_mm}
  \hat\varrho_{\rm sep}=\iint\!\! d\vec\alpha d\vec\beta\,{\mathcal P}(\vec\alpha,\vec\beta)\,\ket{\vec\alpha,\vec\beta;N}\bra{\vec\alpha,\vec\beta;N}.
\end{equation}
We will show that for the separable states (\ref{sep_mm}) and the interferometric transformations (\ref{trans}), the phase estimation sensitivity is bounded by the shot-noise.

The sensitivity of the phase estimation is limited by the Cramer-Rao lower bound \cite{holevo2011probabilistic}
\begin{equation}\label{crlb}
    \Delta\theta\geqslant\frac1{\sqrt m}\frac1{\sqrt{F_q}}.
\end{equation}
Here, $m$ is the number of the independent repetitions of the estimation experiment. The $F_q$ is called the quantum Fisher information (QFI) and it quantifies the amount of information about $\theta$, 
which can be extracted from $m$ measurements using any estimation strategy \cite{braunstein1994statistical}. For pure states, the QFI is simple to calculate,
\begin{equation}\label{pure}
F_q=4(\av{\hat J_n^2}-\av{\hat J_n}^2)\equiv4\av{(\Delta\hat J_n)^2},
\end{equation}
where the mean is calculated with the state undergoing the interferometric transformation. For mixed states, the QFI is much more complex,
\begin{equation}
F_q=2\sum_{i,j}\frac{(p_i-p_j)^2}{p_i+p_j}\left|\bra i\hat J_n\ket j\right|^2,
\end{equation}
where $\ket{i/j}$ are the eigen-vectors and $p_{i/j}$ are the corresponding eigen-values of the density matrix. Therefore, to obtain the $F_q$, one would need to diagonalize $\hat\varrho_{\rm sep}$
from Eq.~(\ref{sep_mm}), which is numerically feasible, but analytically very hard, since different $\ket{\vec\alpha,\vec\beta;N}$'s are not orthogonal, just as nonorthogonal are the coherent states
of light. However, a useful feature of the QFI---its convexity---allows to lower-bound the sensitivity for mixed stated. Namely, for $\hat\varrho=\sum_ip_i\hat\varrho_i$,
i.e., a statistical mixture of density matrices, we have
\begin{equation}
    F_q\left[\sum_ip_i\hat\varrho_i\right]\leqslant\sum_ip_iF_q[\hat\varrho_i].
\end{equation}
This property, applied to Eq.~(\ref{sep_mm}), gives
\begin{equation}\label{convex}
    F_q\leqslant\iint\!\! d\vec\alpha d\vec\beta\,{\mathcal P}(\vec\alpha,\vec\beta)F_q^{(\vec\alpha,\vec\beta)}.
\end{equation}
Here, $F_q^{(\vec\alpha,\vec\beta)}$ is the QFI calculated with a pure state $\ket{\vec\alpha,\vec\beta;N}$, thus it is given
by Eq.~(\ref{pure}).

We calculate the QFI for $\hat J_{\vec n}=\hat J_z$ and show that for separable states, $F_q\leqslant N$. Any other direction $\vec n'$ can be obtained by a series of rotations of $\hat J_z$,
generated by the operators (\ref{mJ}). They are single-particle objects, therefore, once applied to the separable state (\ref{sep_mm}), rather than to the evolution operator, they would
transform one $\hat\varrho_{\rm sep}$ into another, i.e., would only change $\mathcal P(\vec\alpha,\vec\beta)$ into some $\tilde{\mathcal P}(\vec\alpha,\vec\beta)$, but will not modify the structure of
Eq.~(\ref{sep_mm}). Thus it is enough to show that $F_q\leqslant N$ for $\hat J_z$ and a general probability distribution.

First, we calculate $F_q^{(\vec\alpha,\vec\beta)}$ with Eq.~(\ref{pure}). The mean of $\hat J_z$ is
\begin{equation}\label{msq}
  \av{\hat J_z}=\frac N2\sum_n\left(|\alpha_n|^2-|\beta_n|^2\right)\equiv\frac N2(|\vec\alpha|^2-|\vec\beta|^2),
\end{equation}
where we used the expression (\ref{mJz}) and the relation
\begin{subequations}
\begin{eqnarray}
&&\hat a_n\ket{\vec\alpha,\vec\beta;N}=\sqrt N\alpha_n\ket{\vec\alpha,\vec\beta;N-1},\\
&&\hat b_n\ket{\vec\alpha,\vec\beta;N}=\sqrt N\beta_n\ket{\vec\alpha,\vec\beta;N-1}.
\end{eqnarray}
\end{subequations}
The second moment is calculated in a similar way, i.e.,
\begin{equation}
    \av{\hat J_z^2}=\sum_{n,m}\av{\hat J_z^{(n)}\hat J_z^{(m)}}=\sum_{n}\av{\left(\hat J_z^{(n)}\right)^2}+\sum_{n\neq m}\av{\hat J_z^{(n)}\hat J_z^{(m)}}.
\end{equation}
The two contributions must be treated separately
\begin{eqnarray*}
    \sum_n\av{\left(\hat J_z^{(n)}\right)^2}&=&\frac N4\left[1+(N-1)\sum_n(|\alpha_n|^2-|\beta_n|^2)^2\right],\\
    \sum_{n\neq m}\av{\hat J_z^{(n)}\hat J_z^{(m)}}&=&\frac{N(N-1)}4\times\\\
    &\times&\sum_{n\neq m}(|\alpha_n|^2-|\beta_n|^2)(|\alpha_m|^2-|\beta_m|^2).
\end{eqnarray*}
We add these two terms and obtain
\begin{equation}
  \av{\hat J_z^2}=\frac N4+\frac{N(N-1)}4(|\vec\alpha|^2-|\vec\beta|^2)^2.
\end{equation}
The substraction of the squared mean from Eq.~(\ref{msq}) gives
\begin{equation}\label{bound}
  0\leqslant F_q^{(\vec\alpha,\vec\beta)}=N\left(1-(|\vec\alpha|^2-|\vec\beta|^2)^2\right)\leqslant N.
\end{equation}
This result, combined with Eq.~(\ref{convex}) yields
\begin{equation}\label{snl}
  F_q\leqslant N.
\end{equation}
Since the QFI can surpass the SNL, for instance with two-mode spin-squeezed states, we conclude that the particle entanglement is the resource for the SSN metrology also in multi-mode systems.
We now extend the above formalism to distinguishable particles.

\subsection{Distinguishable particles}\label{sec.ent.dist}
\subsubsection{Multi-mode interferometric transformations}

Note that the formalism of second quantization allows for a quick generalization of the above results to distinguishable particles.
Namely, another index must be attributed to the operators $\hat a_n$ and $\hat b_n$, such that labels the species of the particle (and of the associated field). This way, for the particle of type $j$,
we obtain
$\hat a^{(j)}_n$ and $\hat b^{(j)}_n$.
The interferometric transformations cannot transmute a particle of one type into another---such process would violate the conservation laws and the related super-selection rules
\cite{ssr_wick,ssr2,wasak2016role}. Thus a particle of each type undergoes a separate transformation, which means that the operators (\ref{J_mm}) change into
\begin{subequations}\label{op_dist}
  \begin{eqnarray}
    &&\hat J_x=\frac12\sum_{j=1}^N\sum_n\left(\hat a^{(j)\dagger}_n\hat b^{(j)}_n+\hat b^{(j)\dagger}_n\hat a^{(j)}_n\right)\\
    &&\hat J_y=\frac1{2i}\sum_{j=1}^N\sum_n\left(\hat a^{(j)\dagger}_n\hat b^{(j)}_n-\hat b^{(j)\dagger}_n\hat a^{(j)}_n\right)\\
    &&\hat J_z=\frac12\sum_{j=1}^N\sum_n\left(\hat a^{(j)\dagger}_n\hat a^{(j)}_n-\hat b^{(j)\dagger}_n\hat b^{(j)}_n\right).\label{Jz_mm}
  \end{eqnarray}
\end{subequations}
Once the interferometric transformations are determined, we discuss the role of the particle entanglement for the SSN sensitivity. 
\subsubsection{Role of the particle entanglement}
The separable state of $N$ distingishable particles is constructed from the one-body pure states. For the particle of type $j$ distributed among $a$ an $b$ it reads
\begin{equation}\label{css_dist}
  \ket{\vec\alpha^{(j)},\vec\beta^{(j)};N}=\left(\vec\alpha^{(j)}\hat{\vec a}^{(j)\dagger}+\vec\beta^{(j)}\hat{\vec b}^{(j)\dagger}\right)\ket0.
\end{equation}
The parrallel with the coherent state from Eq.~(\ref{css_mm}) is evident and it is even more pronounced when the separable state of distinguishable particles is introduced
\begin{widetext}
  \begin{equation}\label{sep_dist}
    \hat\varrho_{\rm sep}=\iint d\vec\alpha^{(1)}d\vec\beta^{(1)}\ldots\iint d\vec\alpha^{(N)}d\vec\beta^{(N)}\mathcal{P}(\vec\alpha^{(1)},\vec\beta^{(1)},\ldots,\vec\alpha^{(N)},\vec\beta^{(N)})
    \bigotimes_{i=1}^N\ket{\vec\alpha^{(j)},\vec\beta^{(j)};N}\bra{\vec\alpha^{(j)},\vec\beta^{(j)};N}.
  \end{equation}
\end{widetext}
The state of $N$ distinguishable and non-entangled particles is the state (\ref{sep_mm}) but with each mode---now labeled with two numbers $n$ and $j$, rather than with $n$ only---
occupied with one particle. The deep analogy holds since the transformation (\ref{op_dist}) cannot transmute the particles.

Once this close relation is noticed, the QFI can be bounded from above in a similar fashion to that presented in Section~\ref{sec.qfi} and the calculation is straightforward. We again pick the
interferometric transformation to be generated by $\hat J_z$ and using the convexity of the QFI declared in Eq.~(\ref{convex}), we have
\begin{widetext}
  \begin{equation}\label{qfi_dist}
    F_q\leqslant\iint d\vec\alpha^{(1)}d\vec\beta^{(1)}\ldots\iint d\vec\alpha^{(N)}d\vec\beta^{(N)}\mathcal{P}(\vec\alpha^{(1)},\vec\beta^{(1)},\ldots,\vec\alpha^{(N)},\vec\beta^{(N)})
    \sum_{j=1}^NF_q^{(\vec\alpha^{(j)},\vec\beta^{(j)})}.
  \end{equation}
\end{widetext}
Here we used the fact that the operator (\ref{Jz_mm}) acts on each particle independently. Every $F_q^{(\vec\alpha^{(j)},\vec\beta^{(j)})}$ is calculated with a single-particle state
(\ref{css_dist}), therefore it is bounded as in Eq.~(\ref{bound}) but with $N=1$ (because there is only a single particle of a given type), namely
\begin{equation}
  0\leqslant F_q^{(\vec\alpha^{(j)},\vec\beta^{(j)})}=\left[1-(|\vec\alpha^{(j)}|^2-|\vec\beta^{(j)}|^2)^2\right]\leqslant 1.
\end{equation}
Therefore the sum from Eq.~(\ref{qfi_dist}) is bounded as follows
\begin{equation}
  \sum_{j=1}^NF_q^{(\vec\alpha^{(j)},\vec\beta^{(j)})}\leqslant N.
\end{equation}
Since $\mathcal{P}$ from Eq.~(\ref{qfi_dist}) is normalized we obtain
\begin{equation}
  F_q\leqslant N
\end{equation}
for all separable states of distinguishable particles defined in Eq.~(\ref{sep_dist}). On the other hand, take an entangled NOON state of $N$ distinguishable particles---for instance each in the same
spatial mode---in the form
\begin{equation}
  \ket\psi=\frac1{\sqrt2}\left(\prod_{j=1}^N\hat a^{(j)\dagger}+\prod_{j=1}^N\hat b^{(j)\dagger}\right)\ket0.
\end{equation}
This state will give $F_q=N^2$ (the Heisenberg scaling), therefore the QFI is a witness of particle entanglement also for distinguishable particles.

\subsection{Other cases}\label{sec.ent.other}

For the story to be complete we must consider two other possible cases. One is when the system contains both distinguishable particles and bosons, all together forming a separable state.
Again, we repeat that the interferometric transformations---to be consistent with the conservation laws and the super-selection rules---cannot transmute the particles from these two groups into each other.
Therefore, since they form a separable state where the mode occupied by each particle can be addressed individually, the set of distinguishable particles and bosons can be treated separately. Thus to each of
these sets the arguments from the above sections apply, so also in such configuration the particle-entanglement will be a necessary resource to beat the shot-noise limit.

Another possibility is that the system does not contain a fixed number of particles, but rather its amount fluctuates from shot to shot, governed by the probability distribution $P(N)$.
The separable state is now
\begin{equation}
  \hat\varrho_{\rm sep}=\sum_{N=0}^\infty P(N)\hat\varrho^{(N)}_{\rm sep},
\end{equation}
where $\hat\varrho^{(N)}_{\rm sep}$ contains $N$ particles and is given either by Eq.~(\ref{sep_mm}) or (\ref{sep_dist}) or by a mixture of those, as discussed in the above paragraph.
Since the
operators $\hat J_i$ do not couple states with different number of particles, each fixed--$N$ sector can be treated separately. Therefore, using again
the convexity of the QFI, in all the cases we obtain for separable states
\begin{equation}
  F_q\leqslant\sum_{N=0}^\infty P(N)N\equiv\av{N},
\end{equation}
which defines the shot-noise limit \cite{pezze_fluct}.

Finally, we note that these arguments do not apply to collections of fermions, for which the separable states do not exist due to the Pauli exclusion principle.

\section{Estimation from the mean population imbalance}\label{sec.est}

We now abandon the general considerations and switch to a particular phase estimation protocol. We derive the phase sensitivity for a multi-mode Mach-Zehdner interferometer (MZI). 
We take the most common estimation protocol, where the phase is deduced from the average population imbalance
between the two regions. Although the derivation is done for bosons, according to the above arguments the results apply also to
distinguishable particles and collections of both.

In analogy to the two-mode case, the multi-mode MZI evolution operator is
\begin{equation}\label{evo}
    \hat U(\theta)=e^{-i\theta\hat J_y},
\end{equation}
with the generator given by Eq.~(\ref{y_mm}). At the output, in the $i$-th repetition of the experiment, the number of
particles in each arm is measured, i.e., $n_a^{(i)}$ and $n_b^{(i)}$. From this data, the population imbalance between the two sub-systems is calculated, $n_i=\frac{n_a^{(i)}-n_b^{(i)}}2$, and the sequence is repeated $m$ times and averaged to give
\begin{equation}
\overline n_m=\frac1m\sum_{i=1}^mn_i.
\end{equation}
If $m\gg1$, the central limit theorem tells that the probability for obtaining $\overline n_m$ is a Gaussian function
\begin{equation}
p(\overline n_m)\propto\exp\left[-m\frac{\left(\overline n_m-\av{\hat J_z(\theta)}\right)^2}{2\av{(\Delta\hat J_z(\theta))^2}}\right].
\end{equation}
When the phase is estimated from the maximum of the likelihood function for this probability, we obtain
\begin{equation}\label{sens_mzi}
\Delta^2\theta=\frac1m\frac{\av{(\Delta\hat J_z(\theta))^2}}{\left(\frac{\partial\av{\hat J_z(\theta)}}{\partial\theta}\right)^2}.
\end{equation}
The $\hat J_z(\theta)$ is obtained by evolving the $\hat J_z$ with the operator (\ref{evo}), i.e,
\begin{equation}\label{Jz_evo}
    \hat J_z(\theta)=\hat U^\dagger(\theta)\hat J_z\hat U(\theta)=\hat J_z\cos\theta+\hat J_x\sin\theta,
\end{equation}
where the last equality was obtained using the commutation relations of operators (\ref{J_mm}) and the Baker--Campbell--Hausdorff formula.
Substitution of the result (\ref{Jz_evo}) into Eq.~(\ref{sens_mzi}) gives for  $\theta=0$
\begin{equation}
\Delta^2\theta=\frac1m\frac{\av{(\Delta\hat J_z)^2}}{\av{\hat J_x}^2}.
\end{equation}
This is a natural extensions of the spin-squeezing parameter (for balanced systems, where $\av{\hat J_y}=0)$, to the multi-mode case. The combination of Equations (\ref{crlb}) and (\ref{snl})
means that also
\begin{equation}\label{ss_mm}
  \xi^2=N\frac{\av{(\Delta\hat J_z)^2}}{\av{\hat J_x}^2}
\end{equation}
is a witness of quantum correlations among the particles---when $\xi^2<1$ the system is particle-entangled.
We now show that the measurement of this quantity in the multi-mode system is not as straightforward as in the two-mode case.

\section{Detection of entanglement---fluctuations and visibility}\label{sec.det}
First however we recall that for the two-mode systems, the spin-squeezing parameter is defined as
\begin{equation}
  \xi_S^2=N\frac{\av{(\Delta\hat J_3)^2}}{\av{\hat J_1}^2+\av{\hat J_2}^2}.
\end{equation}
Here $1,2,3$ are three orthogonal directions obtained by the combinations of $\hat J_x$, $\hat J_y$ and $\hat J_z$ from Equations (\ref{J_mm}).
\begin{figure}[t!]
  \includegraphics[clip, scale=0.25]{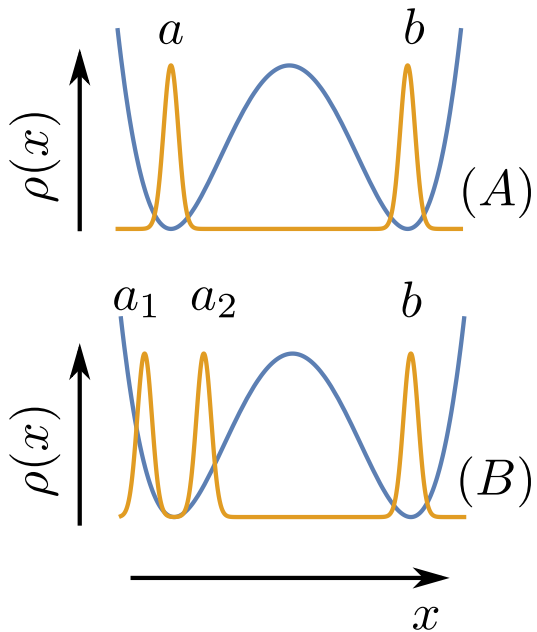}
  \caption{The single-particle density of atoms trapped in a double-well potential. (A) The standard two-mode setup. (B) The three-mode setup:
    the configuration after a coherent splitting of the $a$ mode into $a_1$ and $a_2$.}
  \label{fig.dw}
\end{figure}
When $\xi_S^2<1$, the system is spin-squeezed---it is particle-entangled and useful for quantum metrology \cite{giovannetti2004quantum}.
It it most common to take $(1,2,3)$ as $(x,y,z)$, giving
\begin{equation}\label{ss}
  \xi_S^2=N\frac{\av{(\Delta\hat J_z)^2}}{\av{\hat J_x}^2+\av{\hat J_y}^2}.
\end{equation}
The motivation for this particular choice lies in direct experimental accessibility both to the nominator and the denominator of Eq.~(\ref{ss}).
This can be particularly easily seen when the average number of atoms in each mode is high. In this case, the mode operators can be approximated as follows
\begin{equation}\label{mf}
  \hat a\simeq\sqrt{n_{a}}e^{i\frac\phi2},\ \ \ \hat b\simeq\sqrt{n_{b}}e^{-i\frac\phi2}.
\end{equation}
Here, $n_{a/b}$ are the average mode occupations and $\phi$ is the relative phase, which fluctuates from shot to shot.
In this approximation, the average of $\hat J_x$ and $\hat J_y$ is
\begin{equation}
    \av{\hat J_x}=\sqrt{n_an_b}\av{\cos\phi},\ \ \ \av{\hat J_y}=\sqrt{n_an_b}\av{\sin\phi}.
\end{equation}
Similarly, the atom number fluctuations normalized to the shot-noise is
\begin{equation}\label{fluct}
  \xi^2_N=N\frac{\av{(\Delta\hat J_z)^2}}{n_an_b}.
\end{equation}
Therefore, the spin squeezing from Eq.~(\ref{ss}) can be expressed as follows
\begin{equation}\label{eff}
  \xi_S^2=\frac{\xi_N^2}{\av{\cos\phi}^2+\av{\sin\phi}^2}.
\end{equation}
Note that for the spin-coherent state $\xi_N^2\equiv1$, while the phase is fixed and does not fluctuate from shot to shot, so that
$\av{\cos\phi}^2+\av{\sin\phi}^2=\cos^2\phi+\sin^2\phi=1$, giving $\xi_S^2=1$: the system is not spin-squeezed.
In the experiments, the atom number fluctuations are calculated by measuring in each shot the population of each mode.
In another series of experiments, the two modes are let interfere and the relative phase of the pattern is recorded to give, after many runs, the denominator of Eq.~(\ref{eff}).

Note that there is an alternative method for measuring (\ref{ss}), which does not rely on the mean-field approximation (\ref{mf}).
In the far-field, the spatial wave-packet of the modes $a$ and $b$ are plane-waves (or, more precisely broad functions with fast oscillations on top), i.e.,
\begin{equation}
  \hat\Psi(x)=e^{ikx}\hat a+e^{-ikx}\hat b.
\end{equation}
The density of this system is
\begin{eqnarray}
  \rho(x)&=&\av{\hat\Psi^\dagger(x)\hat\Psi(x)}=\av{\hat a^\dagger\hat a}+\av{\hat b^\dagger\hat b}+\nonumber\\
  &+&2\av{\hat J_x}\cos(2kx)+2\av{\hat J_y}\sin(2kx).
\end{eqnarray}
Since $\av{\hat a^\dagger\hat a}+\av{\hat b^\dagger\hat b}=N$, the normalized density reads
\begin{eqnarray}\label{normalized}
  p(x)&=&\frac1N\rho(x)=1+\nu_1\cos(2kx)+\nu_2\sin(2kx)=\nonumber\\
  &=&1+\nu\cos(2kx-\alpha),
\end{eqnarray}
where $\nu_{1/2}=\frac2N\av{\hat J_{x/y}}$ and $\alpha=\arccos\left(\frac{\nu_1}\nu\right)$, while $\nu=\sqrt{\nu_1^2+\nu_2^2}$ is the fringe visibility.
Therefore, the spin-squeezing (\ref{ss}) can be expressed in terms of the atom number fluctuations $\eta^2=\frac4N\av{(\Delta\hat J_z)^2}$ and the visibility
\begin{equation}\label{op_def}
  \xi_S^2=\frac{\eta^2}{\nu^2}.
\end{equation}
We now demonstrate that the use of this operational definition of the spin-squeezing, without the {\it a priori} knowledge about the mode structure of each arm, can lead to false conclusions regarding the 
presence of the particle entanglement in the system.
\begin{figure}[t!]
  \includegraphics[clip, scale=0.25]{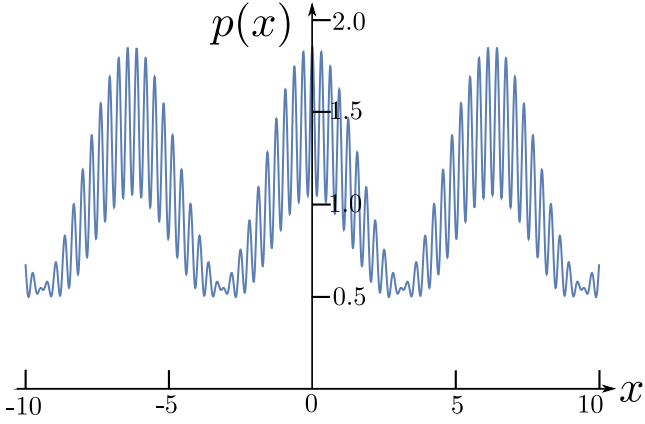}
  \caption{The single-particle far-field interference pattern, see Eq.~(\ref{prob}), formed by a coherent spin-state (\ref{css2}) in the three-mode configuration shown in Fig.~\ref{fig.dw}(B).
  The parameters are $z=0.91$,  $\zeta=0.5$, $k=10$ and $\delta k=0.5$.}
  \label{fig.pattern}
\end{figure}

In the multi-mode case, the density calculated with the operator (\ref{field}) is
\begin{equation}
  \rho(x)=\rho_{aa}(x)+\rho_{bb}(x)+\rho_{ab}(x)+\rho_{ba}(x),
\end{equation}
where $\rho_{ij}(x)=\av{\hat\Psi^\dagger_i(x)\hat\Psi^\dagger_j(x)}$. Clearly the density contains multiple interference terms, not only resulting from the overlap of the wave-functions coming from the opposite
regions but also from different modes residing initially in the same region. Therefore, the visibility of fringes cannot be linked to the denominator of Eq.~(\ref{ss_mm}), which only quantifies the
$a/b$ coherence. For illustration, take a coherent spin state from Eq.~(\ref{css_2m}) with $\varphi=0$. Assume now that in the region $a$, some process splits the mode $a$ into the coherent superposition
of $a_1$ and $a_2$, schematically shown in Fig.~\ref{fig.dw} in a double-well setup. This way, a state
\begin{equation}\label{css2}
  \ket\psi=\frac1{\sqrt{N!}}\Big[\sqrt z\left(\sqrt{\zeta}\hat a_1^\dagger+\sqrt{1-\zeta}\hat a_2^\dagger\right)+\sqrt{1-z}\hat b_1^\dagger\Big]^N\ket0
\end{equation}
is obtained. It is an example of a non-entangled state from Eq.~(\ref{css_mm}) with $\vec\alpha=\sqrt z(\sqrt\zeta,\sqrt{1-\zeta},0\ldots)^{\rm T}$ and $\vec\beta=(\sqrt{1-z},0\ldots)^{\rm T}$. The splitting
does not influence the nominator of Eq.~(\ref{ss_mm}), but in the far-field the interference of the two modes $a_1$ and $a_2$ will have impact on the visibility of fringes. The field operator after
the expansion will read
\begin{equation}
  \hat\Psi(x)=e^{i(k+\delta k)x}\hat a_1+e^{i(k-\delta k)x}\hat a_2+e^{-ikx}\hat b.
\end{equation}
giving the normalized density
\begin{widetext}
  \begin{equation}\label{prob}
    p(x)=\frac1N\av{\hat\Psi^\dagger(x)\hat\Psi(x)}=1+z\sqrt{\zeta(1-\zeta)}\cos(2\delta kx)+\sqrt{z(1-z)}\left(\sqrt\zeta\cos((2k+\delta k)x)+\sqrt{1-\zeta}\cos((2k-\delta k)x)\right).
  \end{equation}
\end{widetext}
This density is plotted in Fig.~(\ref{fig.pattern}) with $z=0.91$,  $\zeta=0.5$, $k=10$ and $\delta k=0.5$. The fringe visibility, calculated as the ratio $\nu=(p_{\rm max}-p_{\rm min})/(p_{\rm max}+p_{\rm min})$
is $\nu^2=0.326$. This combined with the normalized population imbalance between the two regions, $\eta^2=1-(2z-1)^2=0.321$, gives the ratio $\eta^2/\nu^2=0.988$, suggesting the presence of the particle
entanglement in the system. However, it is a false statement: $\nu$ cannot be identified with the denominator of Eq.~(\ref{ss_mm}) in this case.

{\it Some remarks}---Naturally, the three-mode example invoked here is quite artificial. It is hard to imagine that a coherent physical process, which splits $\hat a$ into $\hat a_1$ and $\hat a_2$ could
be uncontrolled and unnoticed by the experimentalists.
Also, any experimentalist would immediately notice two frequencies of oscillations in the interference pattern. Last but not least, usually the multi-mode structure of the two regions $a$ and $b$
comes from the thermal excitations, and thus reveals no coherence between the modes. Nevertheless, the example shows that the method of detecting the particle-entanglement through the
fluctuations-to-visibility ratio can be safely used only when the structure of each sub-system is known.

\section{Summary}\label{sec.sum}

The main outcome of this work is the establishment of the SNL for the two-arm multi-mode interferometers. This is a general result, as it applies to any thinkable quantum system where the entangled/non-entangled dichotomy exists. It is valid for both fixed- and non-fixed-$N$ systems, for as long as the coherence between states carrying different number of particles is absent.

Note that our results are valid for a particular choice of the interferometric transformations, such that do not penetrate the inner structure of each arm but rather treat them as a whole. In principle, any other type of such transformation requires a dedicated calculation of the SNL. Otherwise, conclusions about the presence of metrologically useful particle entanglement in the system can be incorrect. 

We have also calculated the phase sensitivity for the standard estimation protocol based on the knowledge of the mean population imbalance between the two arms. 
Finally, we have shown that if the popular method of detecting the entanglement in two-mode systems is used without the {\it a priori } knowledge about the modal structure of the arms, 
a false conclusion from the number-fluctuations-to-visibility ratio can drawn.

\end{document}